\documentclass[conference,compsoc]{IEEEtran}
\ifCLASSOPTIONcompsoc
  \usepackage[nocompress]{cite}
\else
  \usepackage{cite}
\fi
\ifCLASSINFOpdf
\else
\fi

\usepackage{graphicx}
\usepackage{times}
\usepackage{soul}
\usepackage{url}
\usepackage[hidelinks]{hyperref}
\usepackage[utf8]{inputenc}
\usepackage[small]{caption}
\usepackage{graphicx}
\usepackage{amsmath}
\usepackage{booktabs}
\usepackage{algorithm}
\usepackage{algorithmic}
\usepackage{tikz}
\newcommand*{\circled}[1]{\lower.7ex\hbox{\tikz\draw (0pt, 0pt)%
		circle (.5em) node {\makebox[1em][c]{\small #1}};}}
\usepackage{bbding}
\usepackage{latexsym,bm,amsmath,amssymb}
\usepackage{threeparttable}
\usepackage{subfigure}
\usepackage{makecell}
\usepackage{multirow}
\usepackage{graphicx}
\usepackage{color}
\usepackage{enumerate}
\usepackage{pifont}
\usepackage{caption}
\usepackage{stmaryrd}
\usepackage{todonotes}

\DeclareMathOperator*{\argmin}{arg\,min}

\newtheorem{lemma}{Lemma}

\def\ie{\textit{i.e.}}

\def\eg{\textit{e.g.}}

\begin{document}
%
\title{Shielding Federated Learning: Mitigating Byzantine Attacks with Less Constraints}




\author{\IEEEauthorblockN{Minghui Li\IEEEauthorrefmark{1},
Wei Wan\IEEEauthorrefmark{1},
Jianrong Lu\IEEEauthorrefmark{1}, 
Shengshan Hu\IEEEauthorrefmark{1},
Junyu Shi\IEEEauthorrefmark{1},\\
Leo Yu Zhang\IEEEauthorrefmark{2},
Man Zhou\IEEEauthorrefmark{1}, and
Yifeng Zheng\IEEEauthorrefmark{3}
}
\IEEEauthorblockA{\IEEEauthorrefmark{1}Huazhong University of Science and Technology, Wuhan, China}
\IEEEauthorblockA{\IEEEauthorrefmark{2}Deakin University, Melbourne, Australia}
\IEEEauthorblockA{\IEEEauthorrefmark{3}Harbin Institute of
Technology, Shenzhen, China}
}


%


\maketitle

\begin{abstract}
Federated learning is a newly emerging distributed learning framework that facilitates the collaborative training of a shared global model among distributed participants with their privacy preserved. However, federated learning systems are vulnerable to Byzantine attacks from malicious participants, who can upload carefully crafted local model updates to degrade the quality of the global model and even leave a backdoor. While this problem has received significant attention recently, current defensive schemes heavily rely on various assumptions, such as a fixed Byzantine model, availability of participants' local data, minority attackers, IID data distribution, etc. 

To relax those constraints, this paper presents Robust-FL, the first prediction-based Byzantine-robust federated learning scheme where none of the assumptions is leveraged. The core idea of the Robust-FL is exploiting historical global model to construct an estimator based on which the local models will be filtered through similarity detection. We then cluster local models to adaptively adjust the acceptable differences between the local models and the estimator such that Byzantine users can be identified. Extensive experiments over different datasets show that our approach achieves the following advantages simultaneously: (i) independence of participants' local data, (ii) tolerance of majority attackers, (iii) generalization to variable Byzantine model.

\end{abstract}



%
\IEEEpeerreviewmaketitle

\section{Introduction}

Recently emerged federated learning (FL)~\cite{fl1} is a new computing paradigm that trains a global machine learning model over distributed data while protecting participants’ privacy. By distributing the model learning process to participants, FL constructs a global model from user-specific local models, such that participants' data never leaves their own devices. In this way, the bandwidth cost is significantly reduced and user privacy is well protected. 

Due to the decentralized nature, FL is vulnerable to Byzantine attacks~\cite{SurveyFL,WeightAttack}, where malicious participants can falsify real models or gradients to damage the learning process, or directly poison the training data to make the global model learn wrong information or even leave a backdoor. In the literature, various attack methods have been proposed to demonstrate the vulnerabilities of FL. For example, pixel-pattern backdoor attack~\cite{BadNets} adds a pre-defined pixel pattern to a fraction of training data and modifies the corresponding labels. Label flipping attack~\cite{Sybils} will train the local model by combining correct samples with flipped labels.  These two attacks aim at reducing the recognition rate of the local models by tampering with the training data.
Another kind of attack method focuses on manipulating the local models, such as bit-flip attack~\cite{Zeno} which modifies a part of the local model parameters by flipping specified bits, and sign-flipping attack~\cite{AutoEncoder} which flips the signs of local model parameters and enlarges the magnitudes. Recently a distributed backdoor attack~\cite{xie2019dba} is proposed to show the possibility of uniting multiple participants to conduct an attack, where a backdoor trigger can be decomposed and embedded into different adversarial parties.

To mitigate Byzantine attacks, a mounting number of defense schemes have been proposed~\cite{Krum,FABA,Sniper,SLSGD,DiverseFL}. They mainly focused on comparing participants' local models to remove anomalous ones before aggregating them. These solutions, however, suffer from various limitations that make them unsuitable to be applied in practice. For example, the famous defense scheme Multi-Krum~\cite{Krum} assumes that data is independently and identically distributed (IID) and cannot deal with Non-IID datasets. FABA~\cite{FABA} assumes a fixed Byzantine model and needs to know the number of malicious participants in advance before detection. DiverseFL~\cite{DiverseFL} requires a part of participants' local dataset to help detect anomalous models, which apparently violates the privacy principle of FL.
The most recently proposed defense FLTrust~\cite{FLTrust} is not able to identify the Byzantine participants.
A comprehensive comparison among existing defensive schemes is shown in Table~\ref{Table:1}.

To get rid of these limitations, we propose  Robust-FL, the first prediction-based Byzantine-robust FL scheme. Different from existing works that focus on making use of local models in the current iteration, Robust-FL aims to construct an estimator based on the historical global models from previous rounds.  The local models that significantly differ from the current estimator are expected to have a higher possibility of being malicious, and will be discarded. 
In detail, we first make use of exponential smoothing to construct the estimator, which enjoys a high efficiency for detection, especially when there are large-scale clients in federated learning. We then propose using a small public dataset (\ie, less than 10 samples) to train an initial global model, which is crucial for improving the detection accuracy.

In summary, we make the following contributions:

\begin{itemize}
\item We propose a new Byzantine-robust federated learning scheme called Robust-FL. To the best of our knowledge, Robust-FL is the first predication-based defense scheme that can mitigate Byzantine attacks effectively and efficiently without relying on any {\color{black}fallacious} assumptions.

\item We propose incorporating clustering algorithms to adaptively adjust the differences between the estimator and local models, such that a boundary between benign and malicious models can be effectively affirmed to identify Byzantine participants.

\item We conduct extensive experiments to evaluate Robust-FL. 
The results show that Robust-FL is still effective even more than 50\% participants are compromised, the Byzantine models are variable, and the participants' data are not available, while all existing defenses are invalid under this severe scenario.
\end{itemize}

\section{Related Work}
\label{sec:Related work}
In order to resist Byzantine attacks, researchers have proposed many defensive schemes in recent years. We divide them into three categories according to the principles that the server relied on to detect or evade anomalous local models. 

\textbf{Distance-based defenses:} The first category focuses on comparing the distances between the local models to find out anomalous ones.  Krum~\cite{Krum} aims to choose one local model that is closest to its $K-f-2$ neighbors, where $K$ is the number of participants and $f$ is the number of malicious users. Since Krum converges slowly, the authors introduced its variant Multi-Krum, which chooses $K-f$ local models for aggregation rather than just one. Similar to Multi-Krum, FABA~\cite{FABA} iteratively removes the local model that is farthest from the average model until the number of eliminated models is $f$. FoolsGold~\cite{Sybils} uses cosine similarity to identify malicious models and then assigns them smaller weights to reduce their impact on the averaged global model. Sniper~\cite{Sniper}  selects local models for aggregation based on a graph which is constructed according to the Euclidean distances between the  local models. The PCA scheme~\cite{PCA} projects local updates into two-dimensional space and uses a clustering algorithm to find malicious updates. {\color{black}MAB-RFL~\cite{MAB-RFL} is also equipped with PCA and clustering algorithm to identify malicious updates, in addition, a momentum based approach is applied to tackle the data heterogeneity (\ie, Non-IID) challenge.} All these solutions {\color{black}(except MAB-RFL)}, however, only work well over independently and identically distributed (IID) data, and they cannot tolerate more than 50\% attackers. Besides, most of them need to know the number of attackers in advance.

\textbf{Statistics-based defenses:} The second category exploits the statistical characteristics to remove statistical outliners. Instead of performing detection-then-aggregation, Trimmed Mean~\cite{TrimmedMean} directly uses all the local updates to obtain a new global model, by computing the median or the trimmed mean of all local models in each dimension. Geometric Median~\cite{GeoMed} intends to find a new update that minimizes the summation of the distances between the update and each local model. 
The RFA scheme~\cite{RFA} computes the geometric median of the local models with an alternating minimization approach to reduce the computational overhead. Bulyan~\cite{Bulyan} first uses Multi-Krum to remove malicious models and then aggregates the rest models based on Trimmed Mean. SLSGD~\cite{SLSGD} also adopts Trimmed Mean as the aggregation rule, and then uses a newly proposed moving-average method, which considers global models in this round and the last round. Nevertheless, the above schemes cannot identify Byzantine users, and they perform poorly when there are more than 50\% Byzantine users.

\textbf{Performance-based defenses:} The last category depends on the validation dataset to evaluate the performance of the uploaded parameters. Li et al.~\cite{AutoEncoder} proposed using a pre-trained autoencoder to detect malicious models. Zeno~\cite{Zeno} computes the stochastic descendant score for each gradient based on a validation dataset and then removes the gradients with low stochastic descendant scores. Cao et al.~\cite{NoisyGradient} proposed a Byzantine-robust distributed gradient algorithm, which computes a noisy gradient based on a clean dataset, and a gradient is accepted only when its distance between the noisy gradient satisfies a pre-defined condition. Prakash et al.~\cite{DiverseFL} proposed DiverseFL, which first computes a guiding gradient for each user based on the data the user shares, and then two similarity metrics (Direction Similarity and Length Similarity) between the local gradient and the corresponding guiding gradient are considered, only when both metrics are satisfied will the gradient be accepted. {\color{black} FLTrust~\cite{FLTrust} bootstraps trust with a clean training dataset collected by the server. More specifically, the RELU-clipped cosine-similarity between each local update and the server update (calculated on the cleaning dataset) is employed to reweight the local update, such that malicious updates have a limited impact on the global model.} However, algorithm ~\cite{AutoEncoder} requires a lot of data to obtain benign models and trains autoencoder based on the benign models, but in reality, it is hard to obtain so much data. Although the rest four schemes require few data, they have other limitations. For instance, Zeno needs to know the number of attackers in advances; scheme~\cite{NoisyGradient} relies on an appropriate hyper-parameter to distinguish benign gradients from malicious ones; DiverseFL compels users to share their private data, which violates the original intention of FL; {\color{black}FLTrust cannot identify Byzantine users, which means that malicious updates can also participate in aggregation to deteriorate the accuracy of the global model.}

\begin{table*}[!t]
	\centering
	\renewcommand{\arraystretch}{0.1}
	\caption{A comprehensive comparison among existing defensive schemes. 		\textbf{T(k, d):} the average running time corresponding to the number of users $K$ and the model dimension $d$. Note that for Bulyan $C$ means the time complexity of the aggregation algorithm, and $f$ denotes the number of attackers.
		\textbf{Non-IID Data:} whether the training data is distributed heterogeneously (Non-IID).
		\textbf{50\%-Byzantine:} whether  the percentage of compromised users is larger than 50\%.
		\textbf{N-Independence:} whether  the number of attackers is NOT required in advance.
		\textbf{D-Independence:} whether the shared data that derived from users is NOT required in advance.
		\textbf{Byzantine Identifiability:} whether the Byzantine users can be identified.}\label{Table:1}
 	\resizebox{\textwidth}{!}{
	\begin{tabular}{ |c||c|c|c|c|c|c|c|}
		\hline
		\textbf{Scheme} & \textbf{T(K, d)}&\textbf{Non-IID Data} & \textbf{50$\%$-Byzantine} & \textbf{N-Independence} & \textbf{D-Independence} & \textbf{Byzantine Identifiability}   \\
		\hline
		Multi-Krum~\cite{Krum}  & $\mathcal{O}(K^{2}d)$&{\color{black}\XSolidBrush} & {\color{black}\XSolidBrush} & {\color{black}\XSolidBrush} & {\color{black}\Checkmark} & {\color{black}\Checkmark}  \\
		\hline
		FABA~\cite{FABA}  & $\mathcal{O}(K^{2}d)$&{\color{black}\XSolidBrush} & {\color{black}\XSolidBrush} & {\color{black}\XSolidBrush} & {\color{black}\Checkmark} & {\color{black}\Checkmark} \\
		\hline
		PCA~\cite{PCA}  & $\mathcal{O}(K^{2}d+K^{3})$& {\color{black}\XSolidBrush} & {\color{black}\XSolidBrush} & {\color{black}\Checkmark} & {\color{black}\Checkmark} & {\color{black}\Checkmark}   \\
		\hline
		MAB-RFL~\cite{MAB-RFL}  & $\mathcal{O}(K^{2}d+K^{3})$& {\color{black}\Checkmark} & {\color{black}\XSolidBrush} & {\color{black}\Checkmark} & {\color{black}\Checkmark} & {\color{black}\Checkmark}   \\
		\hline
		Trimmed Mean, Median~\cite{TrimmedMean}  & $\mathcal{O}(Kd \log K)$&{\color{black}\XSolidBrush} & {\color{black}\XSolidBrush} & {\color{black}\Checkmark} & {\color{black}\Checkmark}& {\color{black}\XSolidBrush}  \\
		\hline
		Bulyan~\cite{Bulyan}  & $\mathcal{O}((K-f)C+Kd)$ &{\color{black}\XSolidBrush}& {\color{black}\XSolidBrush} & {\color{black}\XSolidBrush} & {\color{black}\Checkmark}& {\color{black}\Checkmark}  \\
		\hline
		RFA~\cite{RFA}  & $\mathcal{O}(Kd)$&{\color{black}\XSolidBrush} &{\color{black}\XSolidBrush} & {\color{black}\Checkmark} & {\color{black}\Checkmark}& {\color{black}\XSolidBrush}  \\
		\hline
		Sniper~\cite{Sniper}  & $\mathcal{O}(K^{2}d)$&{\color{black}\XSolidBrush} &{\color{black}\XSolidBrush} & {\color{black}\Checkmark} & {\color{black}\Checkmark}& {\color{black}\Checkmark} \\

		\hline
		Resampling~\cite{Resamping}  & $\mathcal{O}(K^{2}d)$& {\color{black}\Checkmark} &{\color{black}\XSolidBrush} & {\color{black}\XSolidBrush} & {\color{black}\Checkmark}& {\color{black}\XSolidBrush} \\
		
		\hline
		
		RSA~\cite{RSA}  & $\mathcal{O}(Kd)$& {\color{black}\Checkmark} & {\color{black}\XSolidBrush} & {\color{black}\Checkmark} & {\color{black}\Checkmark} & {\color{black}\XSolidBrush}  \\
		\hline
		
		DiverseFL~\cite{DiverseFL}  & $\mathcal{O}(Kd)$& {\color{black}\Checkmark} & {\color{black}\Checkmark} & {\color{black}\Checkmark} & {\color{black}\XSolidBrush}& {\color{black}\Checkmark}   \\
		\hline
		FLTrust~\cite{FLTrust} & $\mathcal{O}(Kd)$& {\color{black}\Checkmark} & {\color{black}\Checkmark} & {\color{black}\Checkmark} & {\color{black}\Checkmark}& {\color{black}\XSolidBrush}   \\
				\hline
		Zeno~\cite{Zeno}  & $\mathcal{O}(Kd)$& {\color{black}\Checkmark} & {\color{black}\Checkmark} & {\color{black}\XSolidBrush} & {\color{black}\Checkmark} & {\color{black}\Checkmark}   \\
		\hline
		Our scheme  & $\mathcal{O}(Kd)$& {\color{black}\Checkmark} &{\color{black}\Checkmark} & {\color{black}\Checkmark} & {\color{black}\Checkmark}& {\color{black}\Checkmark}   \\
		\hline
	\end{tabular}
 	}
	\label{tab:compALL}
\end{table*}

\section{Background}
\subsection{Federated Learning}
We consider a general FL system, consisting of a central server and $K$ users. Each user $k$ ($k=1,2,...,K$) has a dataset $D_{k}$, the size of which is denoted as  $|D_{k}|=n_{k}$. It is worth noting that each local dataset may be subject to a different distribution, that is, the users' data may be distributed in a Non-IID way. The users aim to collaboratively train a shared global model $w$. Apparently, the problem can be solved via minimizing the empirical loss, \ie, $\argmin_{w} f(D,w)$, where $D=\bigcup_{k=1}^{K} D_{k}$ and $f(D,w)$ is a loss function (\eg, mean absolute error, cross‐entropy). However, the optimization requires all the users to share their raw data to a central server, which would result in a serious threat to user's privacy. Instead, FL obtains $w$ by optimizing $\argmin_{w} \sum_{k=1}^k f(D_{k},w)$. Specifically, the FL system iteratively performs the following three steps until the global model converges:
\begin{enumerate}[\textbf{Step 1.}]
	\item In the $t$-th iteration, the central server broadcasts a global model $w_{t}$ to the users;
	\item[\textbf{Step 2.}] After receiving $w_{t}$, each user $k$ trains a new local model $w_{t+1}^{k}$ over $D_{k}$ by solving the optimization problem $\argmin_{w_{t+1}^k} f(D_{k},w_{t+1}^k)$ and then uploads it to the server;
	\item[\textbf{Step 3.}] The server aggregates all the local models according to user's proportional dataset size as follows:
	\begin{equation}
	w_{t+1}=\sum_{k=1}^{K} \frac{n_{k}}{n} w_{t+1}^{k}, 
	\end{equation}
	where $n=\sum_{k=1}^{K}n_{k}$. 
\end{enumerate}
\subsection{Exponential Smoothing}\label{sec:back}
Exponential smoothing~\cite{ExpSmo} is a well-known lightweight forecasting algorithm over time series data. It has been widely used in production or economic development forecasting  because of its low computational complexity and high precision. 
Next, we present two definitions that are the cornerstones of exponential smoothing.

\textbf{Definition 1 (First Order Exponential Smoothing)}: 
\textit{If $w_{t}$ is the global model in the $t$-th iteration, and $s_{t-1}^{(1)}$ is the first order exponential smoothing value in the $(t-1)$-th iteration, then the first order exponential smoothing value for iteration $t$ is defined as:
\begin{equation}
s_{t}^{(1)}=\alpha w_{t}+(1-\alpha) s_{t-1}^{(1)},\label{first order exponential smoothing}
\end{equation}
\noindent where $\alpha$ is an empirically determined  parameter ranging between 0 and 1, which adjusts the importance of the latest global models. If $\alpha=1$, then the first order exponential smoothing value is completely determined by the latest global model $w_{t}$, independent of  previous global models. On the contrary, if $\alpha=0$, then the first order exponential smoothing value is independent of global models, and it is always a constant that is determined by the initial value $s_{0}^{(1)}$.}

\textbf{Definition 2 (Second Order Exponential Smoothing)}:
\textit{
	Based on the first order exponential smoothing value, the second order exponential smoothing value in the $t$-th iteration is defined as:
	\begin{equation}
	s_{t}^{(2)}=\alpha s_{t}^{(1)}+(1-\alpha) s_{t-1}^{(2)},\label{second order exponential smoothing}
	\end{equation}
where $s_{t-1}^{(2)}$ is the second order exponential smoothing value in the $(t-1)$-th iteration.  Similarly, we can define $p$-th order exponential smoothing value.}
{\color{black}
\section{Problem Setup}
\subsection{Attack Model}
As is typical in off-the-shelf Byzantine robust defenses~\cite{FLTrust,MAB-RFL,AGRTailored}, we assume that an adversary can control numerous users. It is noteworthy that the adversary may control more than $50\%$ users. However, the FL system must contain at least one benign participant. The adversary can arbitrarily manipulate the training data and the model parameters of the users it controls, corresponding to data poisoning attack and model poisoning attack respectively. Nevertheless, the central server and benign users are not under the control of the adversary. Thus the only way the adversary can distract the global model accuracy is by uploading poisoned local models through the compromised users. Note that the adversary can launch collusion attack, where all malicious models are similar or even identical (\eg, LIE attack~\cite{ALittleIsEnough}), to guarantee the stealthiness of the attack.
\subsection{Defense Model}
Our defense is deployed on the central server and expected to mitigate all kinds of Byzantine attacks with less constrains (as described in Table~\ref{Table:1}). Specifically, the server does not access the raw local training data and is agnostic about the number of compromised users. In addition, the defense should identify Byzantine users accurately so as to reduce the frequencies that these users are selected in further iterations~\cite{MAB-RFL}. Furthermore, the defense is required to tolerate more than $50\%$ attackers and guarantee the quality (\ie, high accuracy) of the global model in both IID and Non-IID scenarios, with a comparable computation overhead with that in FedAvg~\cite{FedAvg}.

Note that the central server is equipped with a small size guiding dataset like many existing defenses~\cite{FLTrust,Zeno,FLOD,Zeno++}. We highlight that, in real-world scenarios, it is toilless for the server to gather such a guiding dataset without sacrificing user privacy (\eg, acquiring from publicly available datasets, manual labeling, or offering voluntarily by users).
}
\section{Robust-FL: Defending Against Byzantine Attacks via Estimation}
\subsection{Key Insight}

After reviewing the existing defenses, we conclude that the main reason behind their limitations is that they focus on making use of the information of the current data in each iteration to detect anomalies, which is indeed a difficult problem. For instance, Multi-Krum tries to remove local models that are far from the overall distribution of the updates uploaded in the current iteration, Zeno  evaluates the performance of each currently uploaded update to discards the bad ones. To this end, those defensive schemes have to make some assumptions to simplify the problem (\eg, consider a fixed Byzantine model) or rely on auxiliary information (\eg, a clean dataset derived from users).

\begin{figure}[t]
	\centering
	\includegraphics[width=\columnwidth]{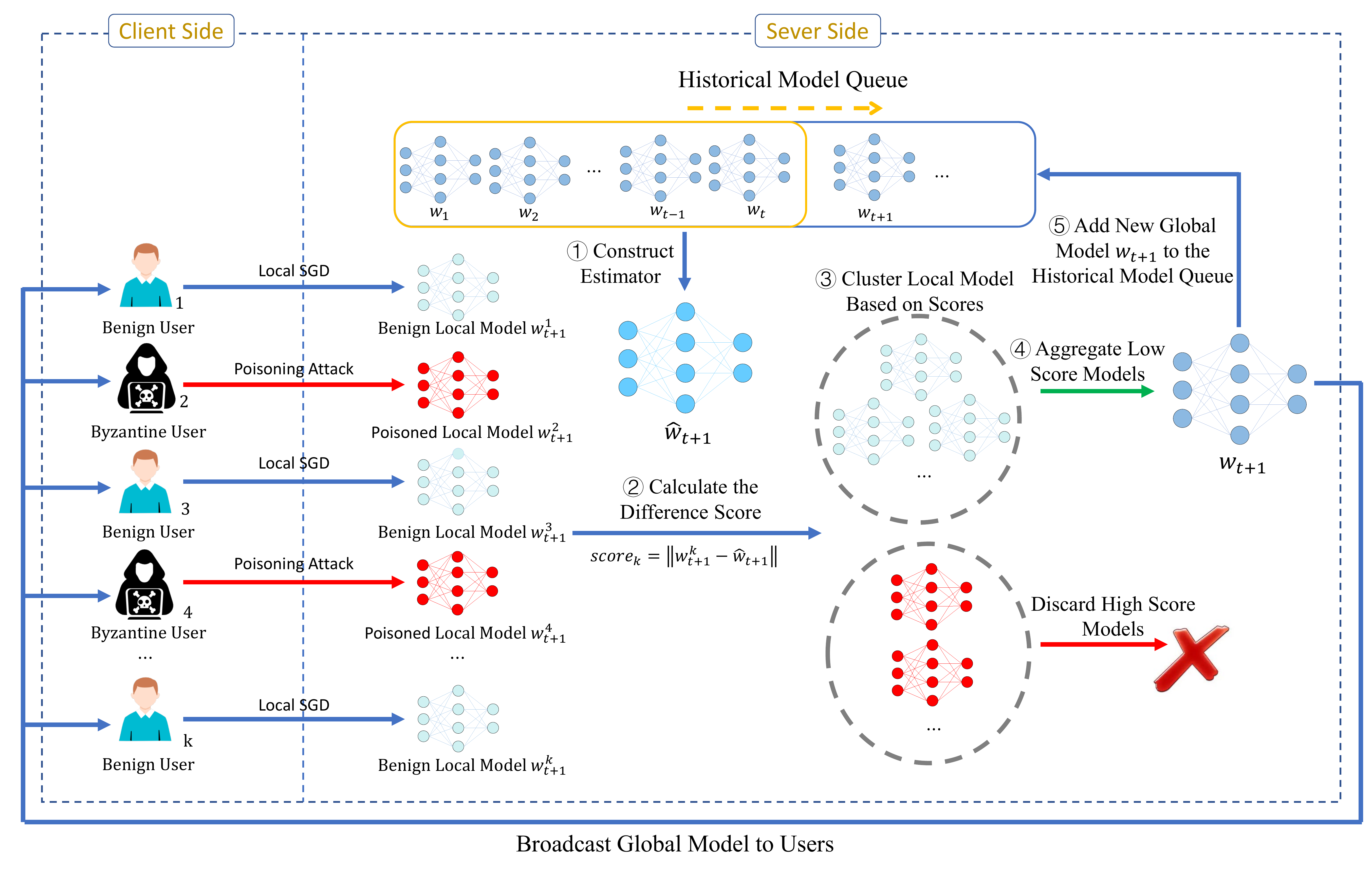}\quad
	\vspace{-2mm}
	\caption{The overview of Robust-FL.}
	\vspace{-4mm}
	\label{fig:OurScheme}
\end{figure}

However, we observe that FL is a typical sequential process and the historical time series data is naturally suitable for predicting and amending future data, as did in recurrent neural networks. In light of this, we aim to make use of the historical data in FL and find that the past global models are of great value in detecting anomalous updates. In brief, in each iteration, Robust-FL first constructs an estimator based on previous global models, and then compares the local models with the estimator. The models that are far away from the estimator will be regarded as malicious and discarded. Then the server aggregates the local models to obtain a new global model and updates the estimator for the next iteration. A brief overview of our scheme is illustrated in Fig.~\ref{fig:OurScheme}. In the next sections, we will show how to construct the estimator by using exponential smoothing and address the two main technical challenges when applying it.

\subsection{Constructing the Estimator}
Robust-FL employs exponential smoothing to construct an estimator, rather than using advanced deep learning algorithms to realize the function of prediction. This is because the server only maintains a sequence of discrete data that is far from enough to train a deep predictive model. Moreover, exponential smoothing enjoys a low computation overhead that enables the server to perform the anomaly detection with high efficiency. 
In this section, we will show how the server computes the estimator $\widehat{w}_{t}$ in the $t$-th iteration.
Based on the exponential smoothing algorithm, we first derive the following lemma:

\begin{lemma}
	\textit{The estimator $\widehat{w}_{t}$ and its $p$-th order exponential smoothing value $s_{t}^{(p)}$ satisfy the following property:
	\begin{equation*}
	\begin{aligned}
	s_{t}^{(p)}=& \sum_{i=0}^{n}(-1)^{i} \frac{\widehat{w}_{t}^{(i)}}{i !} \frac{\alpha^{p}}{(p-1) !} \sum_{j=0}^{\infty} j^{i}(1-\alpha)^{j} \frac{(p-1+j) !}{j !},
	\end{aligned}
	\end{equation*}
	where $\widehat{w}_{t}^{(i)}$ is the $i$-th order derivative of the estimated model $\widehat{w}_{t}$ for $i \in [0, n]$.}
\end{lemma}
Lemma 1 establishes the relationship between $\widehat{w}_{t}^{(i)}$ and $s_{t}^{(p)}$, such that we can obtain $\widehat{w}_{t}^{(i)}$ with $s_{t}^{(p)}$, which are much easier to compute based on Definition 1 and 2 in Section~\ref{sec:back}.

In Robust-FL, we only consider second order exponential smoothing. Let $p=1$ and $p=2$, we can have:
\begin{equation}\label{eq:st1}
\left\{\begin{array}{l}
s_{t}^{(1)}=\widehat{w}_{t}^{(0)} \alpha \sum_{j=0}^{\infty}(1-\alpha)^{j}-\widehat{w}_{t}^{(1)} \alpha \sum_{j=0}^{\infty} j(1-\alpha)^{j}, \\
s_{t}^{(2)}=\widehat{w}_{t}^{(0)} \alpha^{2} \sum_{j=0}^{\infty}(1+j)(1-\alpha)^{j}-\widehat{w}_{t}^{(1)}\alpha^{2} \\
\sum_{j=0}^{\infty} j(j+1)(1-\alpha)^{j}.
\end{array}\right.
\end{equation}

Rearranging Eq.~(\ref{eq:st1}) we can easily obtain:
\begin{equation}\label{eq:wt0}
\left\{\begin{array}{l}
\widehat{w}_{t}^{(0)}=2 s_{t}^{(1)}-s_{t}^{(2)}, \\
\widehat{w}_{t}^{(1)}=\frac{\alpha}{1-\alpha}\left(s_{t}^{(1)}-s_{t}^{(2)}\right).
\end{array}\right.
\end{equation}


On the other hand, according to Taylor series, we can set the estimator $\widehat{w}_{t+T}$ after $T$ iterations as follows:
\begin{equation}\label{eq:wtT}
\widehat{w}_{t+T}=\sum_{i=0}^{n} \frac{\widehat{w}_{t}^{(i)}}{i !} T^{i}.
\end{equation}

In federated  learning, the global model $w_t$ is always adjusted towards the direction of convergence  and the estimator $\widehat{w}_{t}$ only needs to predict the global model for the next iteration, so it is reasonable to assume that $\widehat{w}_{t+T}$ in Eq.~(\ref{eq:wtT}) is linear, in other words, we have
\begin{equation}\label{eq:wt+1}
\widehat{w}_{t+1}=\widehat{w}_{t}^{(0)}+\widehat{w}_{t}^{(1)},
\end{equation}
where we set $T=1$.

Combining Eq.~(\ref{eq:wt0}) and Eq.~(\ref{eq:wt+1}), we have
\begin{equation}\label{eq:wt+1Final}
\widehat{w}_{t+1}=\frac{2-\alpha}{1-\alpha} s_{t}^{(1)}-\frac{1}{1-\alpha} s_{t}^{(2)}.
\end{equation}

In summary, we can use $s_{t}^{(1)}$ and $s_{t}^{(2)}$ to update $\widehat{w}_{t+1}$ easily. Note that it is not required to store all the historical global models on the central server, only $s_{t}^{(1)}$ and $s_{t}^{(2)}$ are needed to obtain the estimator. Therefore, our scheme does not incur additional storage overhead.

\subsection{Initializing Correct Bias Model}

One of the main challenges in Robust-FL is how to generate the initial estimator $w_0$. When using exponential smoothing to detect anomalous local models, a bad $w_0$ will make the estimator converge towards the malicious local models. 

The traditional exponential smoothing usually takes the average of the first several true values as the initial values for $s_{0}^{(1)}$ and    $s_{0}^{(2)}$, and then recursively computes $s_{t}^{(1)}$ and $s_{t}^{(2)}$. However, the solution does not apply to our scheme, because  in the federated learning scenario, there does not exist a true global model at all. 
Another solution is to use a randomly initialized global model $w_{0}$ to compute $s_{0}^{(1)}$ and $s_{0}^{(2)}$. According to Eq.~(\ref{eq:wt+1Final}) the estimator for the first iteration is $\widehat{w}_{1}=w_{0}$, which means that $\widehat{w}_{1}$ is also random. However, the random $\widehat{w}_{1}$ cannot guide the server to accurately identify malicious local models as the server might select a lot of malicious models at first, causing the estimated model to be biased towards malicious models in the subsequent iterations.

In our design, we propose to train an initial global model over a small amount of clean guiding dataset with a few iterations. In Robust-FL, the guiding data can be acquired from any public dataset. In our experiments, we show that only a small amount of public data set, \eg, $10$ samples, can achieve satisfactory accuracy.

\begin{figure}[t]
	\centering
	\subfigure[Evaluation of $\alpha$]{\label{fig:Impact of alpha}
		\includegraphics[width=0.45\columnwidth]{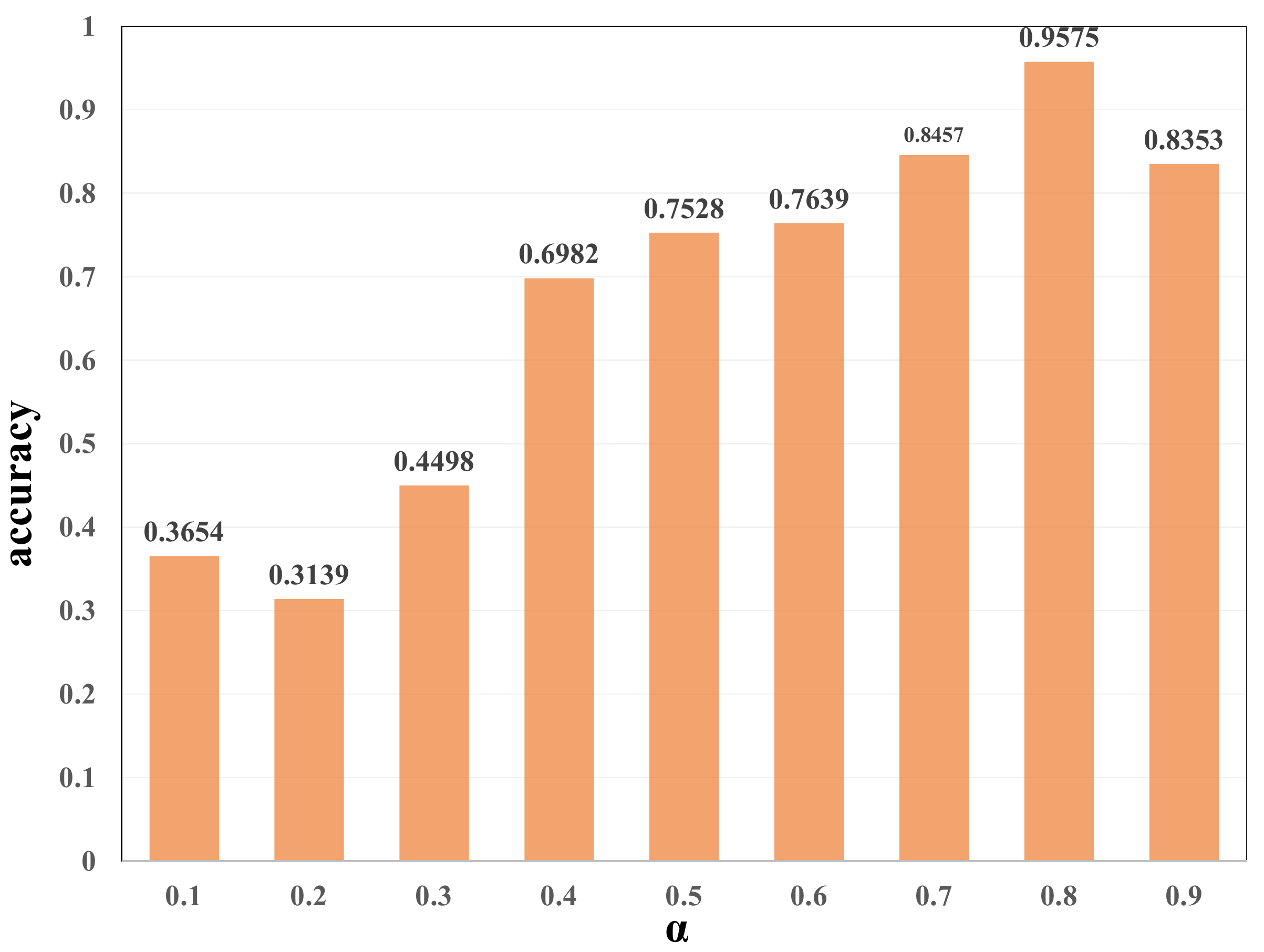}}\quad
	\subfigure[Evaluation of time cost]{
		\includegraphics[width=0.45\columnwidth]{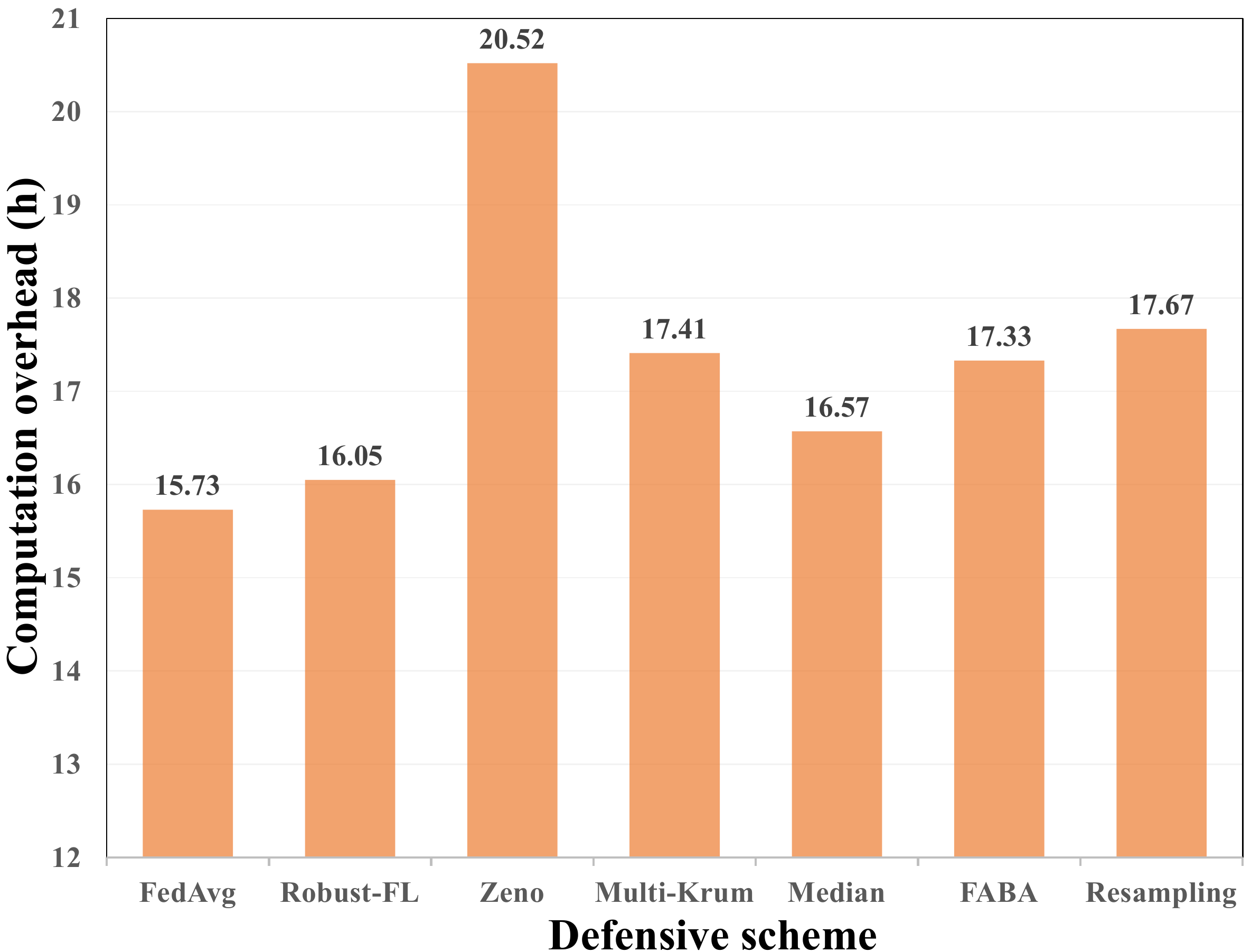}}\quad
	\vspace{-2mm}
	\caption{(a) The impact of $\alpha$ on the accuracy of the global model; (b) The total time overhead after 1000 iterations.}
	\vspace{-4mm}
	\label{fig:alpa&cost}
\end{figure}

\subsection{Identifying Byzantine Users}
Accurately identifying byzantine users and discarding their anomalous local models is of great importance to improve the accuracy of the global model.
However, how to find the boundary between normal and abnormal updates is challenging. 
Most of the existing solutions for identifying Byzantine users rely on the assumption that the Byzantine model is fixed and the number of attackers is known by the server in advance, which makes it much easier to find the boundary. For instance, FABA iteratively eliminates a local model that is farthest from the average model until the number of eliminated models is equal to the number of attackers. Without the assumption, FABA cannot determine how many local models should be eliminated.
\label{section:4.4}


To identify Byzantine users without relying on any assumption, Robust-FL incorporates a clustering algorithm (e.g., $k$-means) based on the bias model. Specifically, we observe that our bias model is able to force the estimator to converge to benign models, making the benign models get much closer to the estimator than malicious ones. Therefore we expect that the biased model will gradually generate a boundary between benign and malicious models. 
In light of this, Robust-FL first calculates the distances between local models and the estimator, and then makes use of $k$-means to categorize them into two classes. The class which has a larger distance with the estimator is regarded as the Byzantine users.

\begin{algorithm}[tb]
	\caption{A Detailed Description of Robust-FL}
	\label{alg:algorithm}
	\textbf{Input}: The local models in $t$-th  iteration  $w_{t}^{1}$, $w_{t}^{2}$, ..., $w_{t}^{K}$; the smoothing factor $\alpha$; the first and second order exponential smoothing values $s_{t}^{(1)}$, $s_{t}^{(2)}$; the public guiding dataset $D_{g}$; the randomly initialized global model $w_{0}$; the number of training iterations over the guiding dataset $T_{g}$.\\
	\textbf{Output}: The global model for the $(t+1)$-th iteration: $w_{t+1}$.
	\begin{algorithmic}[1] 
		\STATE$benign\_model\leftarrow\{\emptyset\}$.
		\IF {$t=0$}
		\STATE $w_{0}=$\textbf{SGD($w_{0}$, $D_{g}$, $T_{g}$)};
		\STATE $s_{0}^{(1)}$=$w_{0}$, $s_{0}^{(2)}$=$w_{0}$.
		\ELSE
		\STATE Construct estimator $\widehat{w}_{t+1}$ using Eq.~(\ref{eq:wt+1Final}).
		\FOR {$k=1,2,3,\cdots,K$}
		\STATE Calculate the difference scores:\\ $
		\text {score}_{t}^{k}=\left\|w_{t}^{k}-\widehat{w}_{t+1}\right\|,
		$
		\ENDFOR
		\STATE Apply $k$-means based on ${score}_{t}^{k}$ to obtain two clusters. Define the class with larger distance with $\widehat{w}_{t+1}$ as $lc$, otherwise as $sc$:\\ $lc$, $sc$=\textbf{KMeans($score_{t}^{1}$,$score_{t}^{2}$,...,$score_{t}^{K}$)}.	
		\FOR{$k=1,2,3,\cdots,K$}
			\IF {$|score_{t}^{k}-sc|<|score_{t}^{k}-lc|$}
			\STATE $benign\_model\leftarrow benign\_model\cup\{w_{t}^{k}\}$.
			\ENDIF
		\ENDFOR
		\STATE$w_{t+1}=FedAvg(benign\_model)$.
		\ENDIF
		\STATE Calculate  $s_{t+1}^{(1)}$ and  $s_{t+1}^{(2)}$ according to definitions in Section \ref{sec:back}.
		\STATE \textbf{return} $w_{t+1}$.
	\end{algorithmic}
\end{algorithm}

\subsection{Robust-FL: A Detailed Illustration}
Algorithm 1 gives a complete description for Robust-FL. Unlike the traditional FL that broadcasts a randomly initialized global model to users, Robust-FL trains the initial model with a small amount of public guiding data to make the estimator be biased towards benign local models (lines 2 to 4). After receiving all the local models, the central server constructs an estimator based on Eq.~(\ref{eq:wt+1Final}) (line 6). Intuitively, benign models will be less different from the estimator compared with malicious ones. So we utilize Euclidean distance (we call it difference score) to measure the differences between each local model and the estimator (lines 7 to 9). Then the $k$-means algorithm is applied to divide the local models into two clusters according to their difference scores (lines 10 to 15).
The cluster with smaller difference scores will be regarded as benign and used for aggregation, while another cluster will be discarded (line 16). Note that  when performing the aggregation, Robust-FL uses  FedAvg~\cite{FedAvg} to save the communication cost and speed up the training process.

\section{Experiments}
\subsection{Experimental Setup}
\emph{1) Datasets and models: }
We use  MNIST and CIFAR-10 to evaluate Robust-FL under different settings. 
MNIST is a 10-class handwritten digit recognition classification dataset contains $60k$ training and $10k$ testing greyscale handwritten digits of size $28 \times 28$.  
CIFAR-10 consists of $50k$ training and $10k$ testing three-channel color images of $10$ different items of size $32 \times 32$. The training samples are  evenly assigned to the users in a random way. 
We train different types of global models on different datasets to show the generality of Robust-FL. Specifically, for MNIST, following previous work~\cite{FedAvg}, we train a convolutional neural network (CNN) as the global model. For CIFAR-10, we use the widely used ResNet20 architecture~\cite{ResNet} as the global model.

\emph{2) Parameter setting: }For MNIST, we set the number of users $K=30$, and consider the increasing percentage of attackers, \ie, $40\%$, $50\%$ and $60\%$. We set the size of guiding dataset $D_{g}$ to $10$. The randomly initialized global model will be trained on $D_{g}$ for $10$ epochs before broadcasting. To reduce the communication overhead, each user trains locally with $3$ iterations to obtain the local model in each epoch. There are $100$ epochs in total.
For CIFAR-10, we set the number of users $K=60$, and consider the percentage of attackers is $30\%$, $40\%$ and $50\%$. The size of $D_{g}$ is set to $20$ and $25$ iterations are required at the beginning. The local iterations and global epochs are set to $3$ and $1,000$ respectively.

\emph{3) Evaluated poisoning attacks: }
In the literature, the poisoning attacks against federated learning can be divided into data poisoning attack and model poisoning attack. For data poisoning attack we consider the popular label-flipping attack where attackers flip their labels from $i$ to $9-i$. For model poisoning attack we evaluate the representative sign-flipping attack where the attackers directly multiply the local model weight with a reverse constant $-c$ to flip the signs of model and adjust its magnitudes. In our experiments, we set a small value $c=0.8$ to reinforce its stealthiness. \textcolor{
black}{Furthermore, we also consider a stronger model poisoning attack LIE (short for ``A little
is enough''~\cite{ALittleIsEnough}), which adds a very small amount of noises to a benign model.}

\begin{figure}[t]
	\centering
	\subfigure[40$\%$ attackers]{
		\includegraphics[width=0.29\columnwidth]{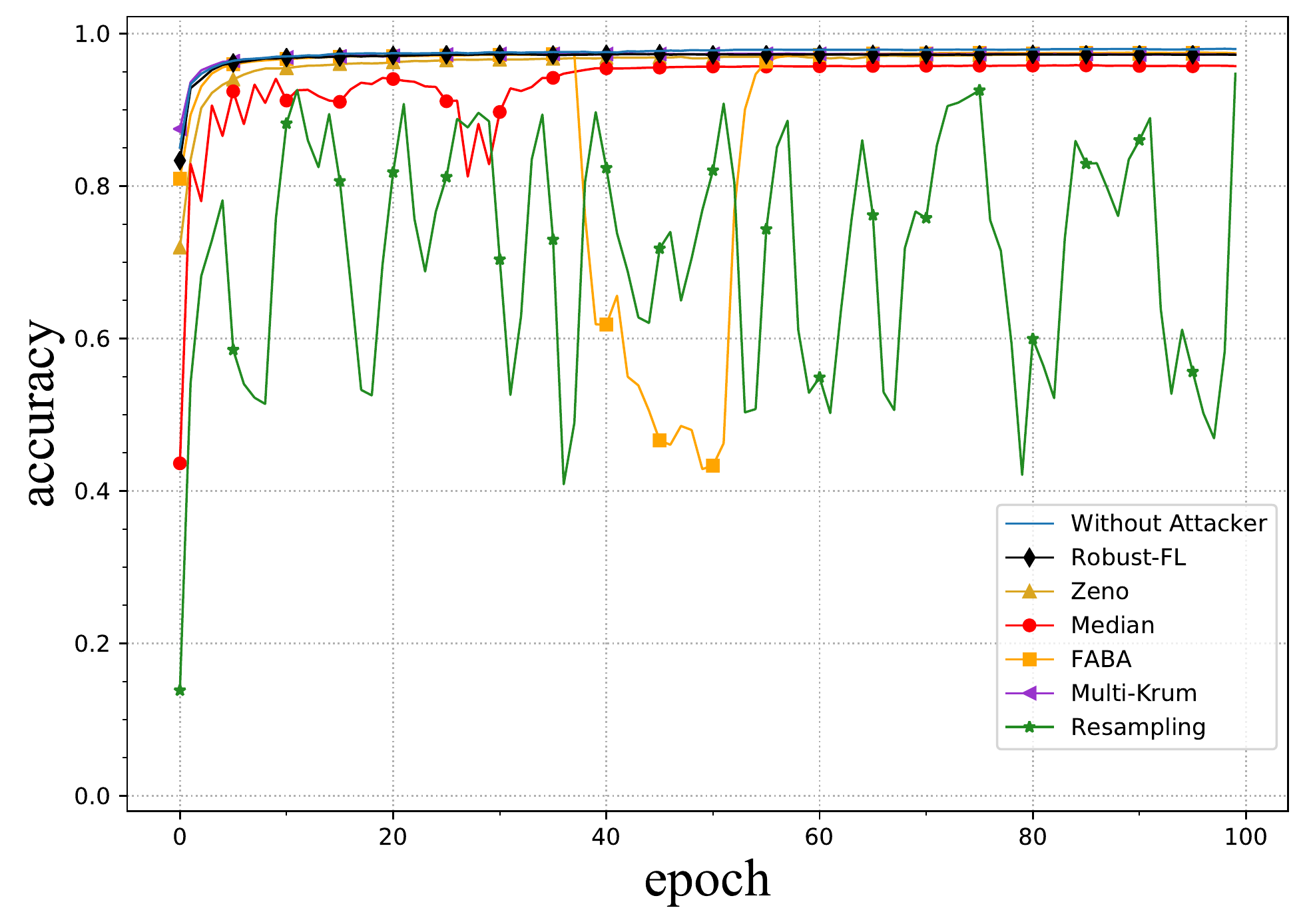}}\quad
	\subfigure[50$\%$ attackers]{
		\includegraphics[width=0.29\columnwidth]{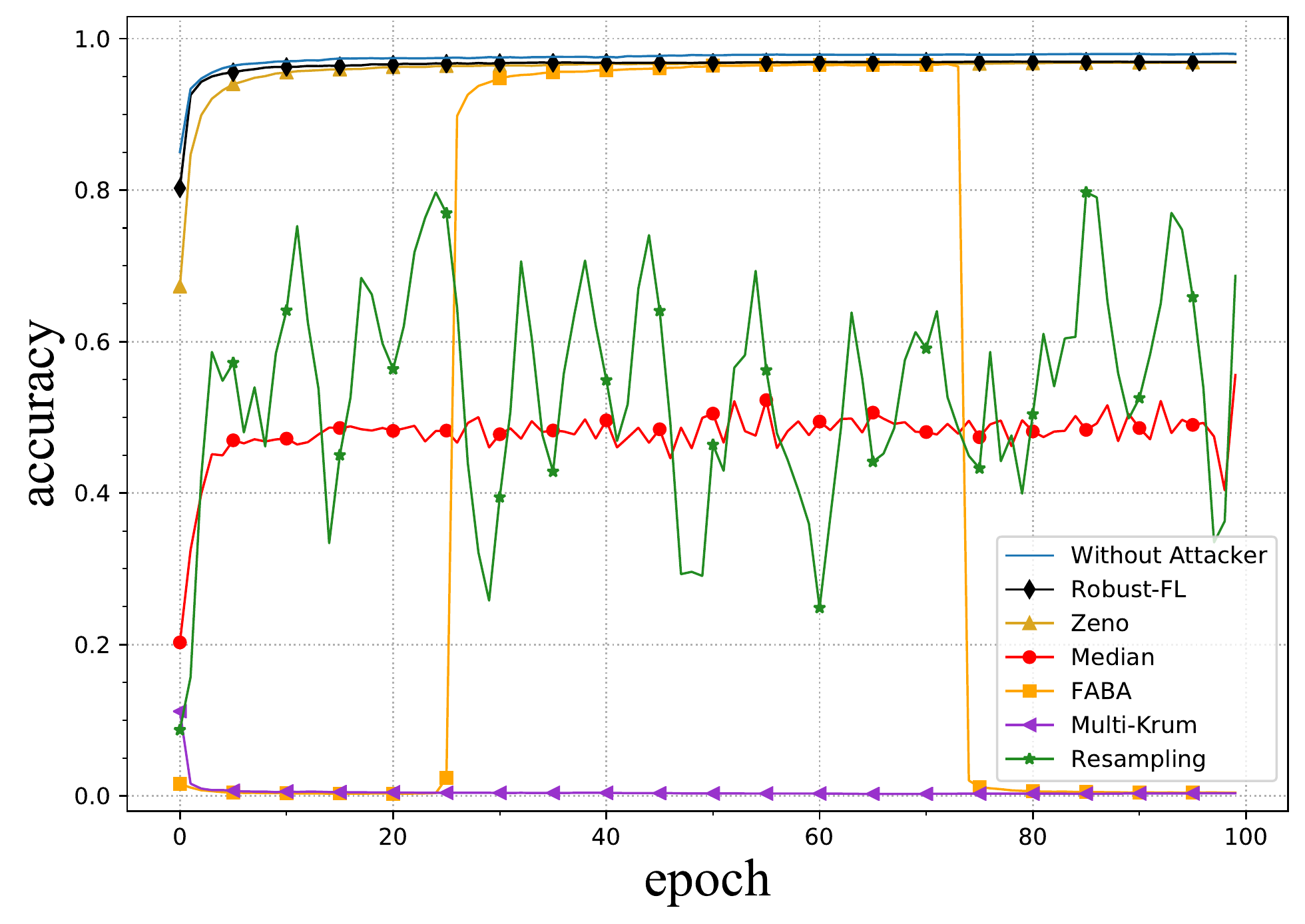}}\quad
	\subfigure[60$\%$ attackers]{
		\includegraphics[width=0.29\columnwidth]{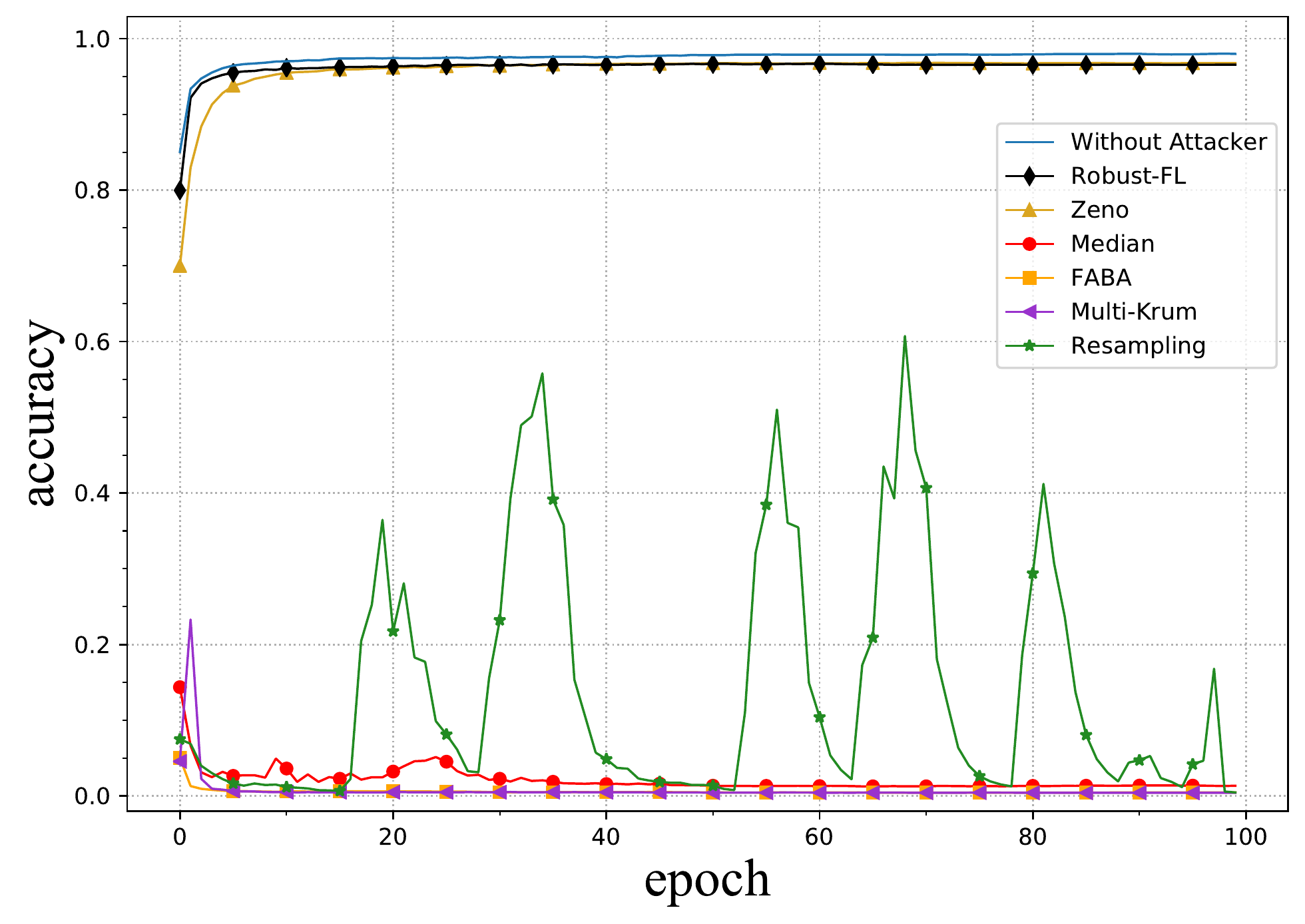}}\quad
	\vspace{-2mm}
	\caption{The accuracy of defensive schemes against label-flipping attack.}
	\vspace{-4mm}
	\label{fig:IID_lf_accuracy}
\end{figure}
\begin{figure}[t]
	\centering
	\subfigure[40$\%$ attackers]{
		\includegraphics[width=0.29\columnwidth]{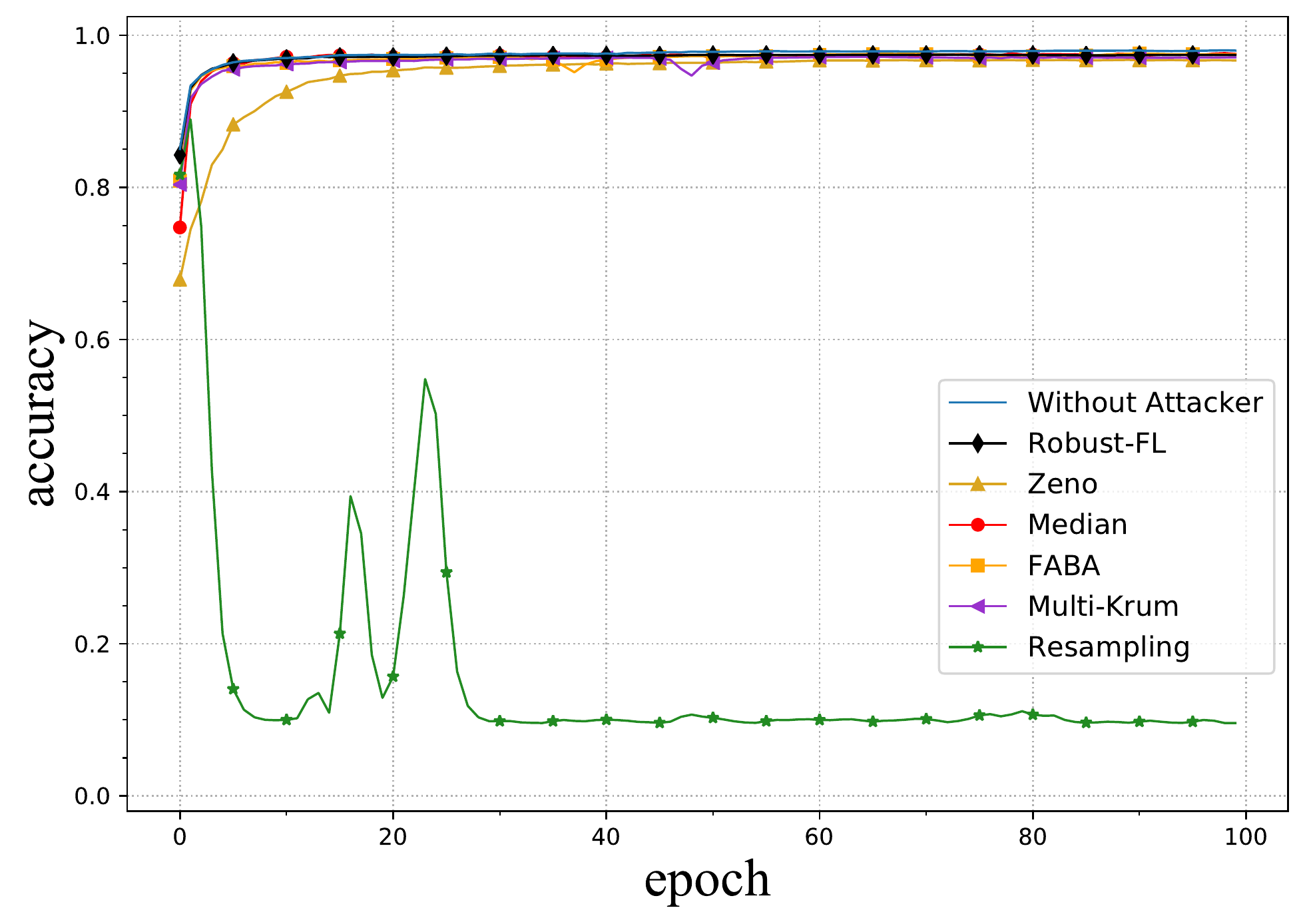}}\quad
	\subfigure[50$\%$ attackers]{
		\includegraphics[width=0.29\columnwidth]{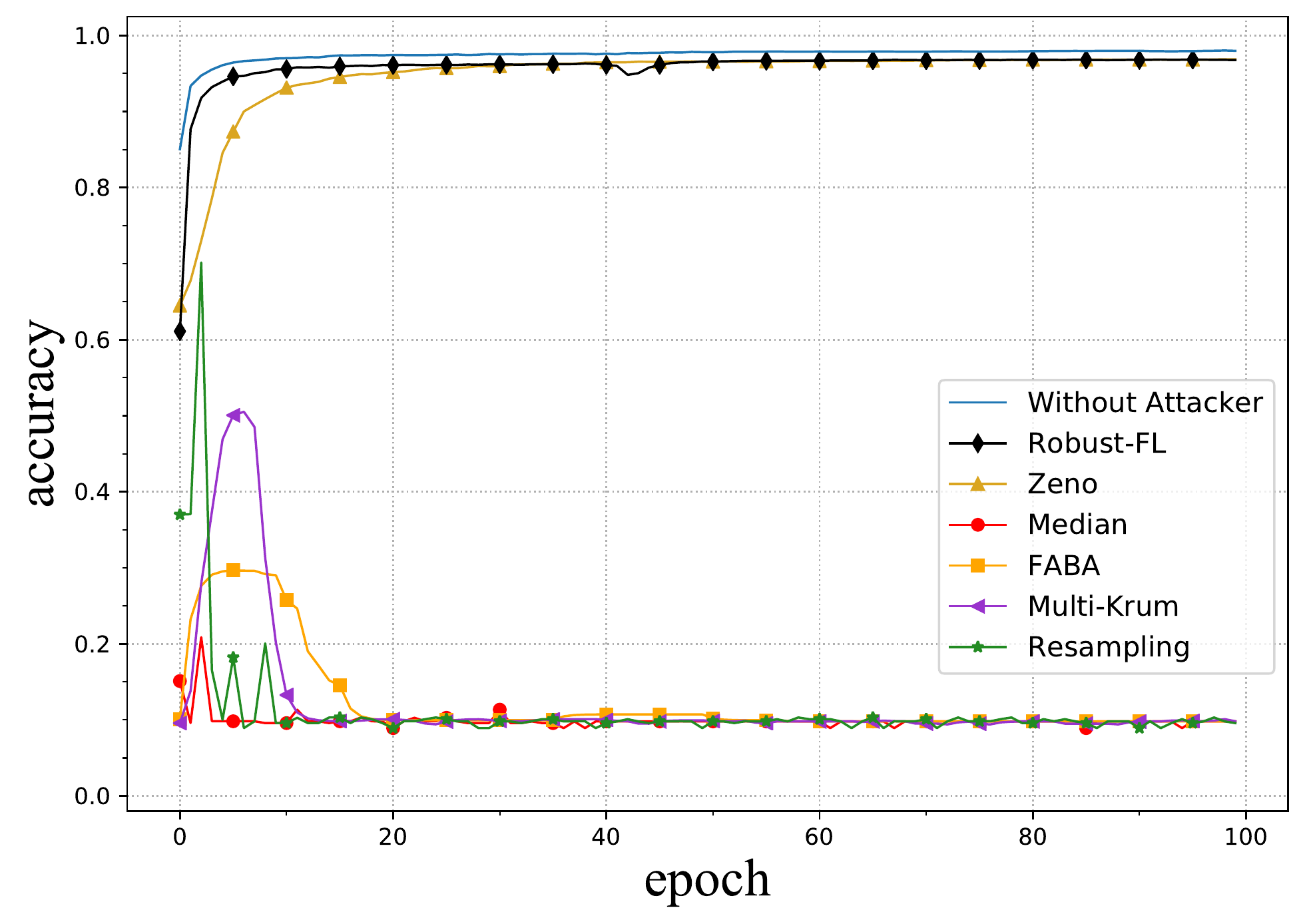}}\quad
	\subfigure[60$\%$ attackers]{
		\includegraphics[width=0.29\columnwidth]{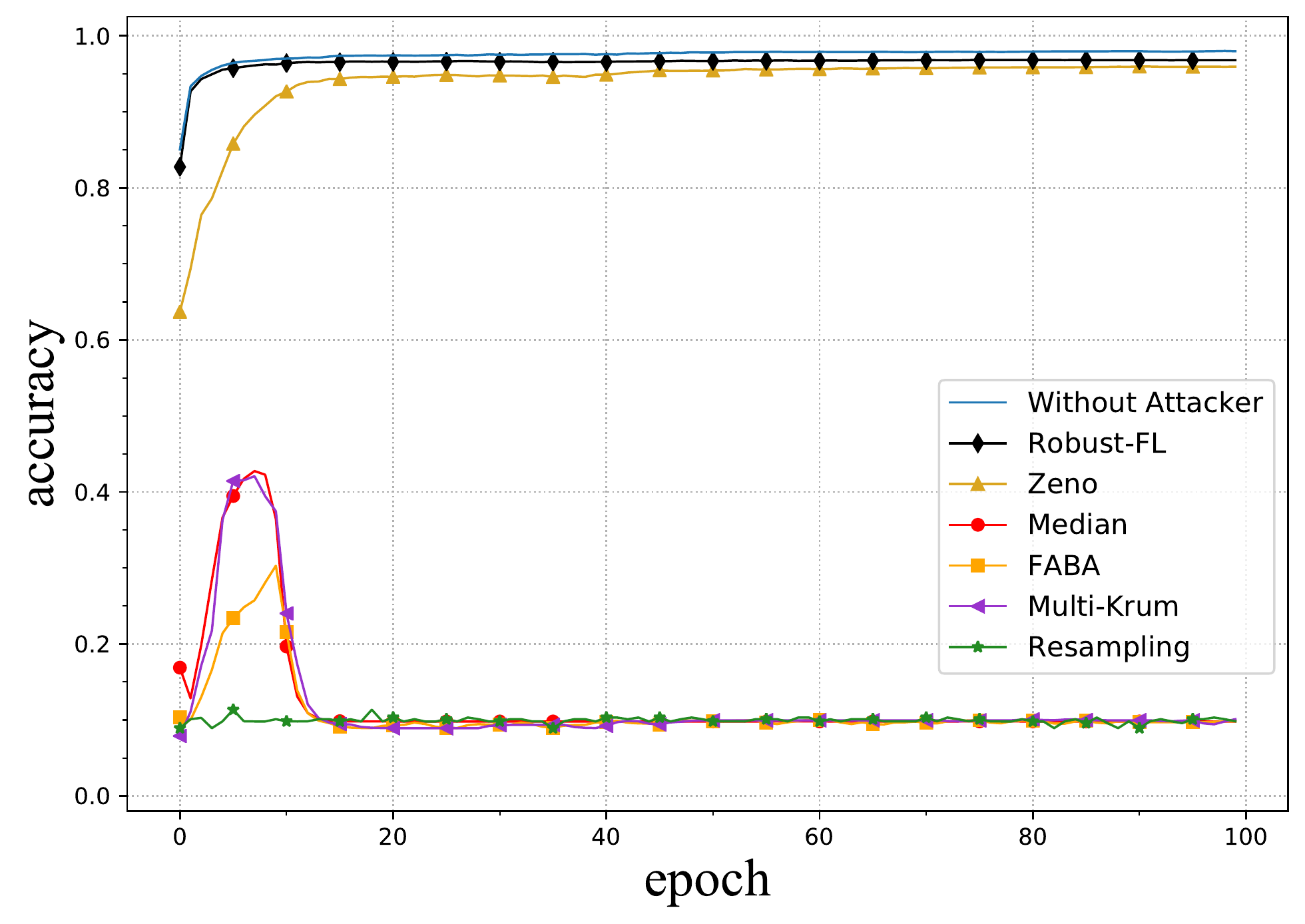}}\quad
	\vspace{-2mm}
	\caption{The accuracy of defensive schemes against sign-flipping attack.}
	\vspace{-4mm}
	\label{fig:IID_sf_accuracy}
\end{figure}

\subsection{Experimental Results}
We compare our proposed Robust-FL with Zeno~\cite{Zeno}, Multi-Krum~\cite{Krum}, FABA~\cite{FABA}, Median~\cite{TrimmedMean} and  Resampling~\cite{Resamping}. In addition, we also implement the baseline where all users are benign. 

\textcolor{black}{
\textbf{Impact of $\alpha$ on global model accuracy:}
In our design,   $\alpha$ determines the portion of the latest global models. It is therefore necessary to figure out the $\alpha$ that provides the best performance. Fig.~\ref{fig:Impact of alpha} shows the accuracy of the global model after $30$ epochs when $\alpha$ varies from $0.1$ to $0.9$. The experiment is conducted on the MNIST dataset where the percentage of label flipping attackers is $50\%$. We can see that the accuracy of global model is lower than $45\%$ when $\alpha$ is smaller than $0.3$. This is because there is a large deviation between the  estimator and the benign local model.  When $\alpha$ increases to  0.8, Robust-FL performs best (with the accuracy of $96\%$). But a larger $\alpha$ does not necessarily indicate a better performance. For example, when $\alpha = 0.9$, the accuracy decreases to $83\%$. We owe this to the fact that the estimator has approached to the global model, and the historical information cannot  be fully utilized, which also leads to the deviation. Therefore we set  $\alpha=0.8$ in our subsequent experiments.
}

\textbf{Computation overhead: }
From Fig.~\ref{fig:alpa&cost}(b), we can see that the computation overhead of Robust-FL is almost the same as that of FedAvg, while the other schemes need much more time to converge, which is consistent with our time complexity analysis in Table \ref{tab:compALL}.
For instance, Robust-FL took 16.05 hours to converge, whereas Zeno, Median, FABA, Multi-Krum and Resampling took 20.52, 16.57, 17.33, 17.41, and 17.67 hours, respectively.  

\textbf{Robustness against label-flipping attack:}
The experimental results over the MNIST dataset in Fig.~\ref{fig:IID_lf_accuracy} show that Robust-FL strengthens the plain federated learning and outperforms state-of-the-art solutions under label-flipping attack. \textcolor{black}{Specifically, Robust-FL achieves about $98\%$ accuracy with minor fluctuations, which implies that Robust-FL has almost the same performance as the baseline (\ie, without attacker).  Zeno performs similar to our scheme. Multi-Krum performs well for $40\%$ attackers, but its performance drops dramatically when the number of malicious users is no less than  $50\%$. FABA and Median perform barely satisfactory in the case of $40\%$ attackers. However, similar to Multi-Krum, these two schemes also perform much worse when the attacker dominants. 
When the percentage of attackers becomes large (\eg, $40\%-60\%$), Resampling fluctuates heavily because it has a high probability to average a new sampling point between the normal models and the malicious models, making it difficult for the central server to decide whether the point should be chosen for aggregation.} 


\textbf{Robustness against sign-flipping attack: }
\textcolor{black}{ Fig.~\ref{fig:IID_sf_accuracy} demonstrates that Robust-FL is resistant to sign-flipping attack. To be specific,  Robust-FL and Zeno perform comparably with the baseline. Multi-Krum, FABA, and Median perform well in the case of $40\%$ attackers, but these schemes become defenceless when the attackers are no less than $50\%$. This is due to the fact that these schemes tend to choose the majority of models with similar behavior for aggregation. Hence the central server is more likely to choose malicious local models when the number of attackers is relatively large. For these reasons, Resmpling performs the worst among the existing defences.}

\begin{figure}[t]
	\centering
	\subfigure[30$\%$ attackers]{
		\includegraphics[width=0.29\columnwidth]{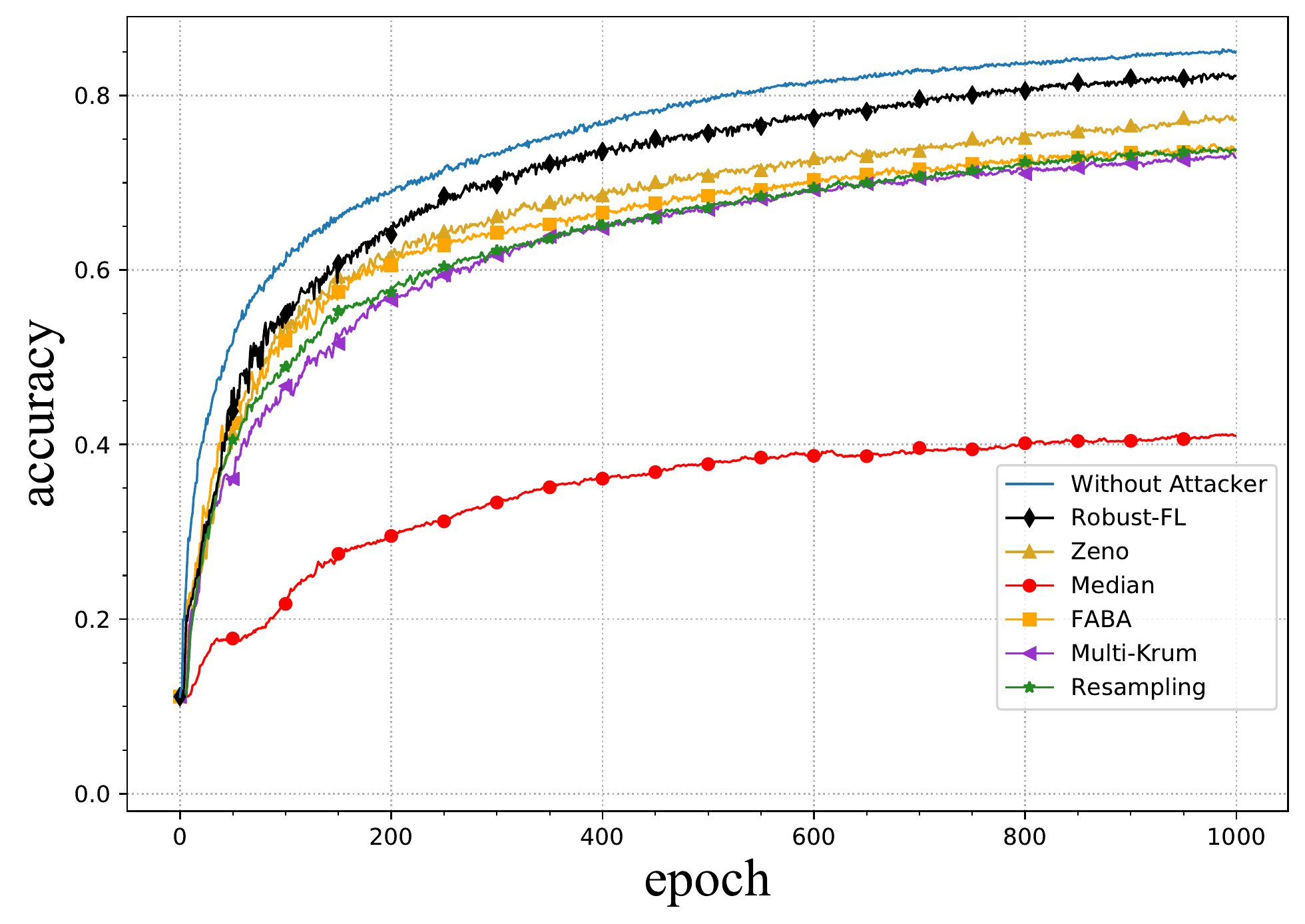}}\quad
	\subfigure[40$\%$ attackers]{
		\includegraphics[width=0.29\columnwidth]{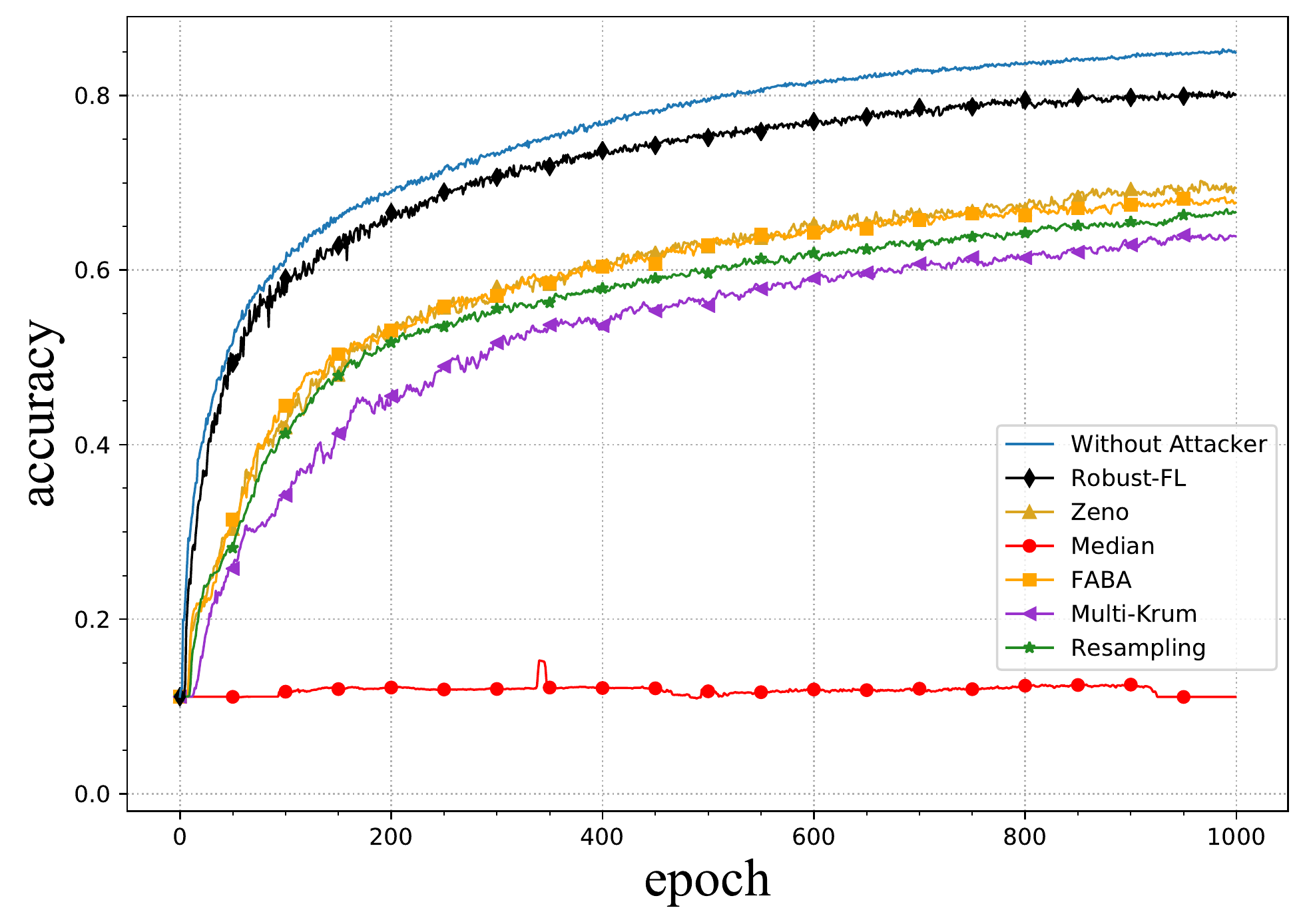}}\quad
	\subfigure[50$\%$ attackers]{
		\includegraphics[width=0.29\columnwidth]{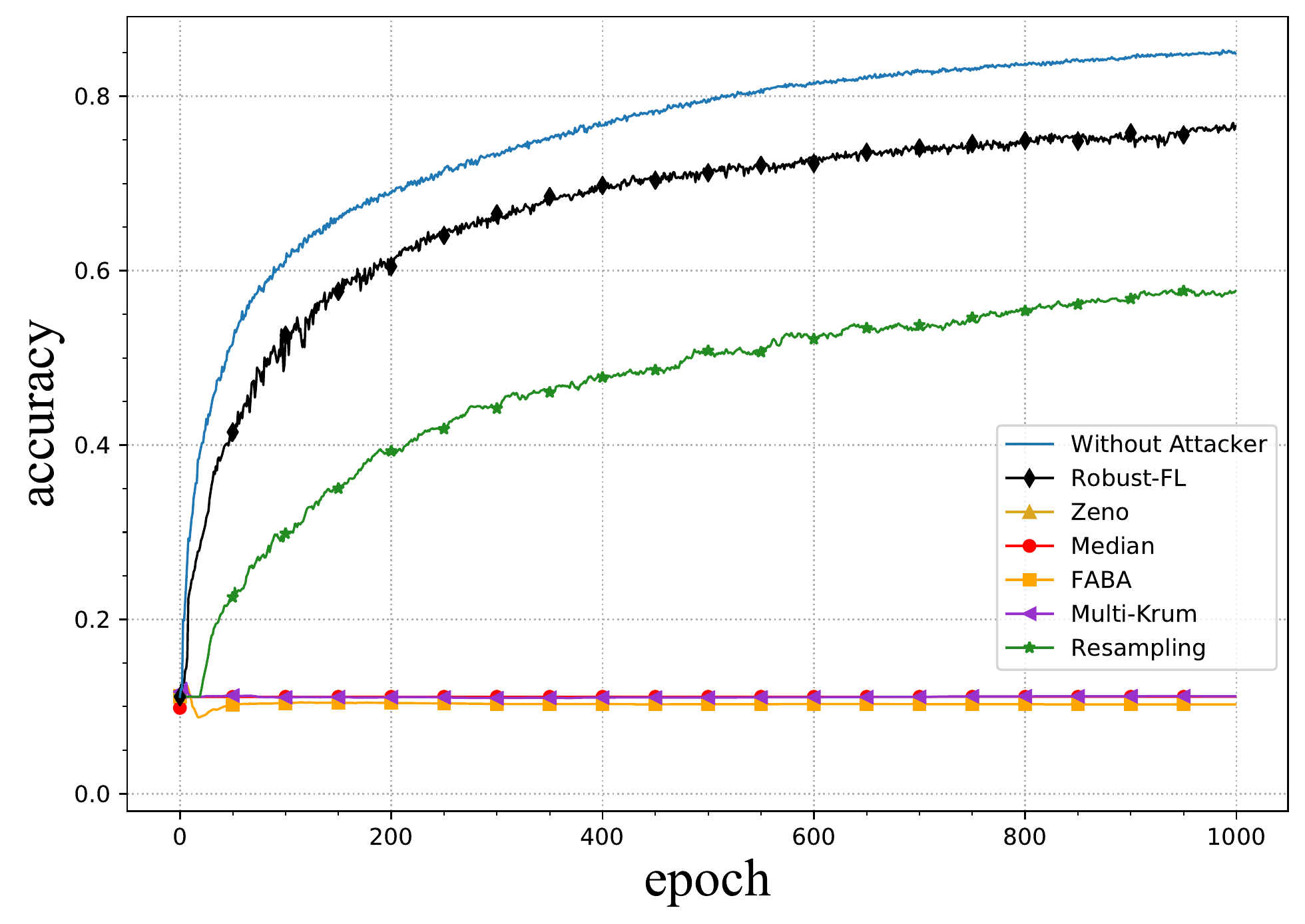}}\quad
	\vspace{-2mm}
	\caption{The accuracy of defensive schemes against LIE attack.}
	\vspace{-4mm}
	\label{fig:IID_lie_accuracy}
\end{figure}

\textcolor{black}{
\textbf{Robustness against LIE attack: } 
Fig.~\ref{fig:IID_lie_accuracy} displays the accuracy of different FL defenses under LIE attack on the CIFAR-10 dataset. Our Robust-FL significantly outperforms the existing solutions. Zeno, FABA, Multi-Krum and Resampling have a similar performance in the case of $30\%$ and $40\%$ attackers, with the accuracy of $10\%-15\%$ lower than Robust-FL. However, when the percentage of attackers is $50\%$, all the schemes become invalid. Under LIE attacks, Median performs the worst all the time. This indicates that LIE attack can circumvent existing defenses by adding a small amount of disturbances, while Robust-FL can effectively resist LIE attack.}


\begin{figure}[t]
	\centering
    	\subfigure[Label-flipping attack]{
		\includegraphics[width=0.45\columnwidth]{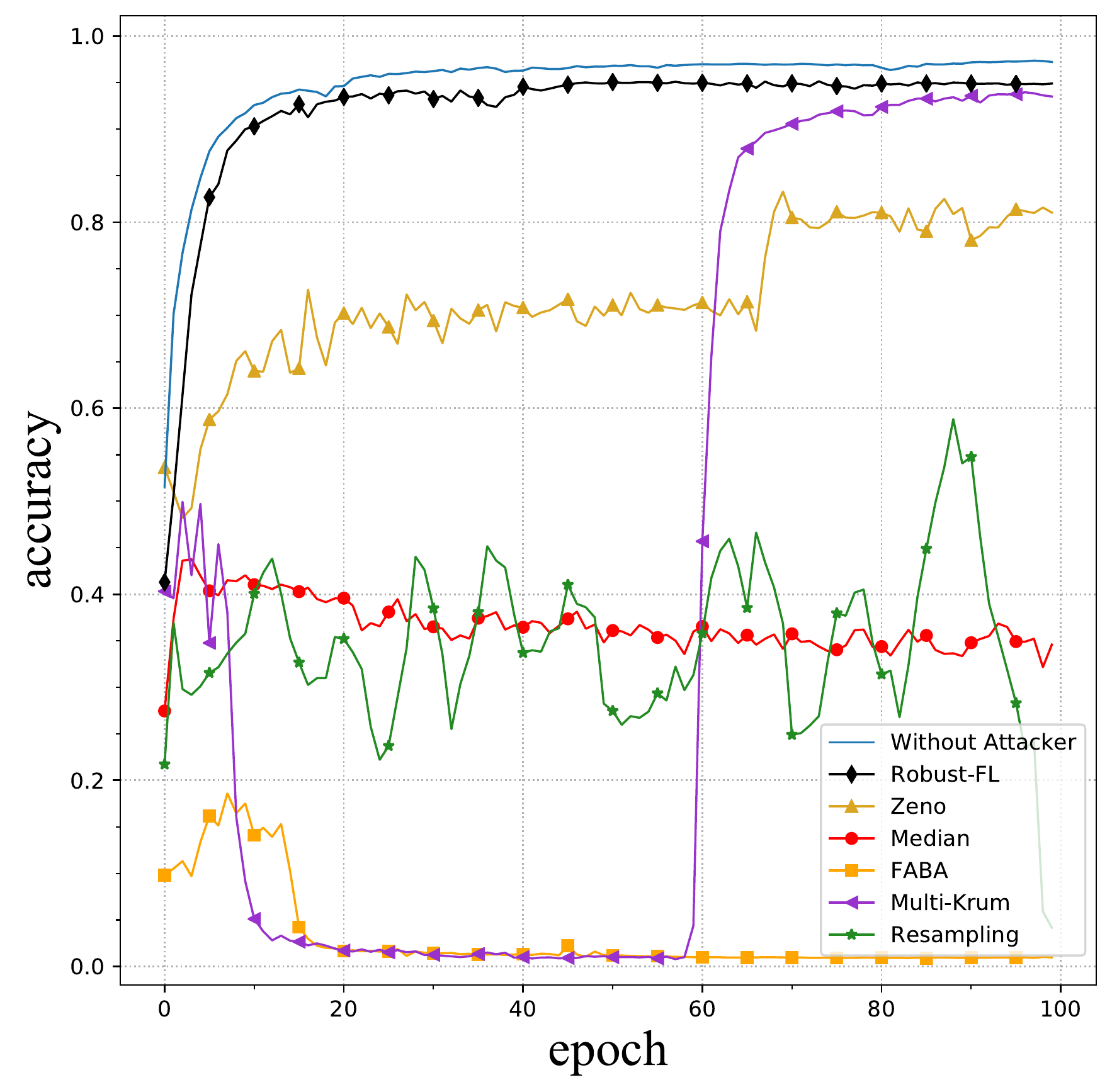}}\quad
    	\subfigure[Sign-flipping attack]{
		\includegraphics[width=0.45\columnwidth]{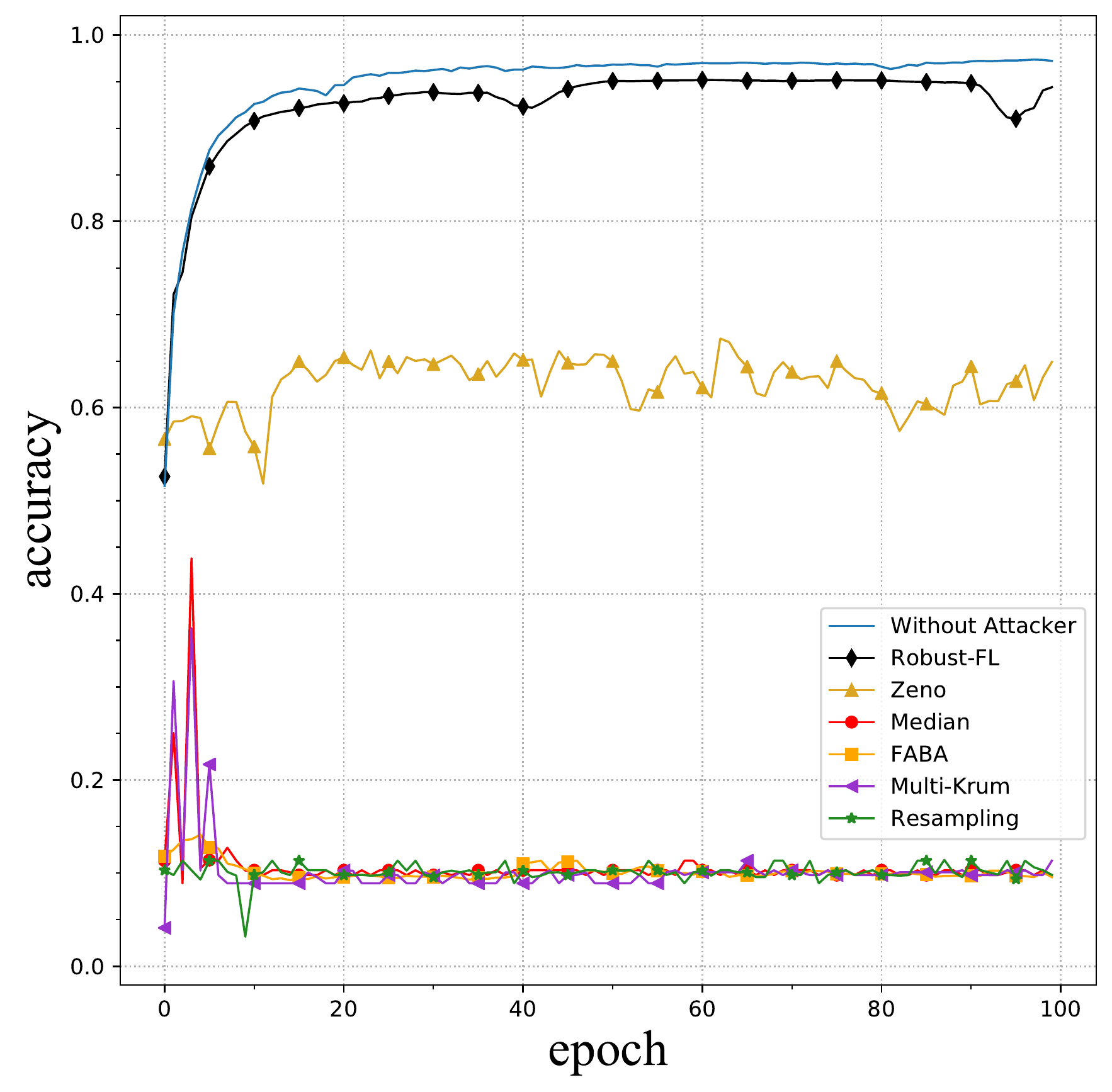}}\quad
		\vspace{-2mm}
    \caption{Evaluations of Robust-FL over Non-IID data.}
    \vspace{-4mm}
	\label{fig:nonIID_accuracy}
\end{figure}

\textbf{Evaluation over Non-IID data:}
Fig.~\ref{fig:nonIID_accuracy} evaluates the performance of Robust-FL on Non-IID local training data. We consider label flipping and sign flipping attacks on MNIST, where the percentage of attackers is $50\%$. We generate the Non-IID data in the same way as~\cite{FLTrust}. Specifically, the Non-IID degree is controlled by a hyper-parameter $q$ between $0$ and $1$. A larger $q$ indicates a higher degree of Non-IID. In the experiments, we consider a strong Non-IID degree where $q=0.95$. We observe that our scheme performs much better than any other defense and is close to the baseline. Zeno, which performs very well in the case of IID setting, has an accuracy of $15\%-30\%$ lower than Robust-FL. The other defenses are completely uncompetitive. 

\section{Conclusion}
This paper focused on defending against Byzantine attacks with relaxed assumptions. We proposed the first estimator-based Byzantine-robust scheme Robust-FL, which constructs an estimator based on the historical global models and then eliminates the  model updates that significantly differ from the estimator. In addition, we utilized clustering algorithms to adjust the acceptable differences between the model updates and estimator adaptively such that Byzantine users can be identified. Experiments on different datasets showed Robust-FL achieved the following advantages simultaneously (i) tolerance of majority attackers, (ii) generalization to variable Byzantine model, (iii) lower computation overhead.

\section*{Acknowledgments}
Minghui’s work is supported in part by the National Natural Science Foundation of China (Grant No 62202186). Shengshan’s work is supported in part by the National Natural Science Foundation of China (Grant Nos. 62002126, U20A20177), and Fundamental Research Funds for the Central Universities (Grant No. 2020kfyXJJS075). Yifeng’s work is supported in part by the Guangdong Basic and Applied Basic Research Foundation (Grant No. 2021A1515110027). Wei's work is supported in part by Ant Group. Shengshan is the corresponding author.

\bibliographystyle{splncs04}
\bibliography{main}

\end{document}